\documentclass[aps,showpacs,pre,twocolumn,superscriptaddress]{revtex4} 

\usepackage{graphicx}% Include figure files
\usepackage{dcolumn}% Align table columns on decimal point
\usepackage{bm}% bold math
\usepackage{epstopdf}

\begin{document}
\title{Slow, Non-Diffusive Dynamics in Concentrated Nanoemulsions}
\author{ H. Guo}
\affiliation{Department of Physics and Astronomy, Johns Hopkins University, Baltimore, Maryland 21218}
\author{J. N. Wilking}
\affiliation{Department of Chemistry and Biochemistry,  University of California-Los Angeles, Los Angeles, California 90095}
\author{D. Liang}
\affiliation{Department of Physics and Astronomy, Johns Hopkins University, Baltimore, Maryland 21218}
\author{T. G. Mason}
\affiliation{Department of Chemistry and Biochemistry,  University of California-Los Angeles, Los Angeles, California 90095}
\affiliation{Department of Physics and Astronomy,  University of California-Los Angeles, Los Angeles, California 90095}
\author{J. L. Harden}
\affiliation{Department of Physics,  University of Ottawa, Ottawa, Ontario K1N 6N5, Canada}
\author{R. L. Leheny}
\affiliation{Department of Physics and Astronomy, Johns Hopkins University, Baltimore, Maryland 21218}

\date{\today}
\begin{abstract}

Using multispeckle x-ray photon correlation spectroscopy, we have measured the slow, wave-vector dependent dynamics of concentrated, disordered nanoemulsions composed of silicone oil droplets in water.  The intermediate scattering function possesses a compressed exponential lineshape and a relaxation time that varies inversely with wave vector.  We interpret these dynamics as strain in response to local stress relaxation.  The motion includes a transient component whose characteristic velocity decays exponentially with time following a mechanical perturbation of the nanoemulsions and a second component whose characteristic velocity is essentially independent of time.  The steady-state characteristic velocity is surprisingly insensitive to droplet volume fraction in the concentrated regime, indicating that the strain motion is only weakly dependent on the droplet-droplet interactions.

\end{abstract}

\maketitle

\section{Introduction}

An emulsion is a dispersion of a liquid within another, immiscible liquid.  In most cases, emulsions are not thermodynamically stable but rather are long-lived, metastable states formed through arrested phase separation or mechanical mixing~\cite{larson}.   In the formation of emulsions through mixing, drops of one liquid are broken apart within another liquid through strong shear flow.  For emulsions comprised of low viscosity liquids, such as oil in water, the shear flows accessible in common mixing devices can produce micrometer-scale droplets that are stabilized against coalescence by added surfactant.  However, with extremely strong shear one can create nanoemulsions containing droplets with diameters well below 100 nm, as recent work employing high-pressure flows has demonstrated~\cite{meleson}.  At low concentrations, the droplets in nanoemulsions behave like spherical colloids that diffuse freely.  At higher concentrations, the droplets enter a jammed state, and the nanoemulsions behave as disordered elastic solids.  However, due to the deformable nature of the droplets, the structure of concentrated nanoemulsions can deviate significantly from that of typical colloidal glasses~\cite{graves,tomJPCB}.  For example, the volume fraction of droplets in concentrated nanoemulsions can exceed the random close packed limit for spheres, implying faceted droplet shapes like in a macroscopic foam.  Thus, in many respects concentrated nanoemulsions represent a new type of nanostructured soft material.  

While the structural properties of concentrated nanoemulsions have recently been elucidated through small angle scattering studies~\cite{graves,tomJPCB}, less is understood about the nature of the droplet dynamics in this glassy state.  To investigate these dynamics, we have performed x-ray photon correlation spectroscopy (XPCS) experiments on a series of concentrated nanoemulsions.  The large wave vectors and long time scales accessible with XPCS make the technique ideally suited for probing the motion of the nanoscale droplets in the concentrated, jammed regime.  We find these dynamics are described by extremely slow ballistic motion that can be modeled in terms of strain in response to localized changes in internal stress.  Such dynamics have been observed in a wide range of disordered soft-solid materials including dilute and concentrated colloidal gels~\cite{cipelletti,chung}, clay suspensions~\cite{bandyopadhyay,bellour}, dense ferrofluids~\cite{robert}, concentrated emulsions of micrometer-scale droplets~\cite{cipelletti-farad}, and a polymer-based sponge phase~\cite{falus}.  In contrast to most previously studied systems, in which these dynamics typically evolve in a manner akin to aging in glasses, the slow dynamics in the nanoemulsions display a steady-state component.  The presence of steady-state strain motion in the nanoemulsions thus suggests a procedure for understanding better these dynamics and their significance by investigating systematically their relationship to other material properties.  As an initial effort, we have characterized the dependence of the dynamics on droplet volume fraction.  Surprisingly, we find the strain dynamics are largely insensitive to changing volume fraction within the concentrated regime despite the large change in macroscopic elastic modulus that accompanies the change.

\section{Experimental Procedures}

The nanoemulsions were prepared following procedures described elsewhere~\cite{meleson} and consisted of poly(dimethyl-siloxane) oil droplets in water stabilized by sodium dodecyl sulfate (SDS).  Two sets of samples were included in the study, one with an average droplet radius $a=$ 46 nm, which we label set A, and one with $a=$ 36 nm, which we label set B.  The volume fraction $\phi$ of oil droplets was varied over the ranges $0.33 < \phi < 0.55$ for A and $0.28 < \phi < 0.45$ for B.  The upper limits of these ranges were set by the volume fractions of the stock emulsions produced in the synthesis.  To obtain lower $\phi$, we diluted the stock emulsions with a surfactant solution identical to the continuous phase:  10 mM SDS in water.  Due to the ionic nature of SDS, the surfaces of the oil droplets are negatively charged, and the droplets interact through a Coulombic repulsion with a Debye screening length that is an appreciable fraction of the droplet radii.   To account for the range of interaction between droplets, an effective droplet volume fraction can be estimated as $\phi_{eff} \approx \phi(1 + h/a)^3$~\cite{tomPRE97}, where $h$ is an effective droplet ``shell'' thickness whose size is estimated from the average inter-droplet separation using an interaction potential for repulsive droplets~\cite{larson,russel,rheo}.  This renormalization of volume fraction leads to fluid-solid transitions in nanoemulsions at $\phi_{eff} \approx 0.58$, near the colloidal glass transition, independent of droplet radius and further leads to a scaling of the shear modulus of concentrated nanoemulsions of different radii onto a master curve as a function of $\phi_{eff}$~\cite{rheo}.  The corresponding ranges of effective volume fractions for the nanoemulsions in the XPCS measurements were $0.62 < \phi_{eff} < 0.82$ for set A and  $0.60 < \phi_{eff} < 0.77$ for set B.~~Based on the rheology studies~\cite{rheo}, nanoemulsions over these ranges of $\phi_{eff}$ behave as elastic solids.  However, the shear modulus over these ranges decreases significantly with decreasing $\phi_{eff}$, by at least of factor of 100 for each set~\cite{rheo}.  Efforts to extend the ranges to even lower $\phi_{eff}$ led to emulsions that were fluid and that had droplet dynamics that were too rapid to observe in the XPCS measurements.

The experiments were performed at sector 8-ID of the Advanced Photon Source using 7.35 keV x-rays.  Details regarding the beam line have been presented elsewhere~\cite{lumma_rsi,lumma_pre}.  In this study, a 20 $\mu$m $\times$ 20 $\mu$m aperture before the sample was employed to create the partially coherent x-ray beam.  The nanoemulsions were loaded into sealable, stainless-steel sample holders with thin polyimide windows for transmission scattering and sample thickness of 0.5 mm.   The loading involved extracting appropriate quantities of nanoemulsion from the glass vials in which they are stored and spreading it into the sample cells with a razor.  The temperature of the samples was held at 25.0 C throughout the measurements.  The scattering intensity was recorded by a direct-illuminated CCD area detector positioned 3.4 m beyond the sample to cover a wave-vector range of 0.05 nm$^{-1} < q <$ 0.39 nm$^{-1}$.  A series of scattering images, with a typical exposure time of 0.5 s, was obtained to determine the ensemble-averaged intensity autocorrelation function $g_2(q,t)$ over the range 3 s $< t <$ 1000 s~\cite{lumma_rsi,lumma_pre}.  The minimum delay time, 3 s, was set by the sum of the exposure time plus the time to download each image from the CCD.  To reduce exposure of the sample to x-rays, a shutter before the sample was closed during the data transfer.  The longest delay time was limited to 1000 s to avoid effects of measurement stability that led to artificial decays in $g_2(q, t)$ at several thousand seconds.  The measurement of $g_2(q,t)$ was repeated over the course of several hours for each sample in order to characterize any evolution of the structure or dynamics of the nanoemulsions.  As an additional part of the protocol to avoid radiation damage to the samples, each measurement of $g_2(q,t)$ was performed with the x-ray beam incident on a new region of the sample.    

\section{Results and Discussion}
\subsection{Scattering Intensity and Dynamic Structure Factor}

Figures 1(a) and 1(b) show the x-ray scattering intensity $I(q)$ for a nanoemulsion from set A with $\phi_{eff} = 0.82$ and one from set B with $\phi_{eff} = 0.60$, respectively, measured at several waiting times $t_w$ since the nanoemulsions were loaded into sample cells.  The measured $I(q)$ are unnormalized; therefore, to compare measurements performed on different regions of the samples, we arbitrarily normalize each set of curves with respect to the intensity at the smallest wave vector.  Consistent with previous studies~\cite{graves,tomJPCB}, the scattering intensities display a pronounced peak near $q = 0.10$ nm$^{-1}$, corresponding to the first peak in the interparticle structure factor, as well as additional correlation peaks at higher wave vector.  The structure displays little change with $t_w$ over the time of the experiments, consistent with the very long shelf life of the nanoemulsion samples.  However, we note that $I(q)$ at the largest $t_w$ in Fig.~1(a) does show some deviation from $I(q)$ at earlier $t_w$.  This deviation represents the largest change in $I(q)$ with $t_w$ observed for any nanoemulsion studied.   

%%%%%%%%%%%Figure 1 Here%%%%%%%%%
\begin{figure}
\includegraphics[scale=1.0]{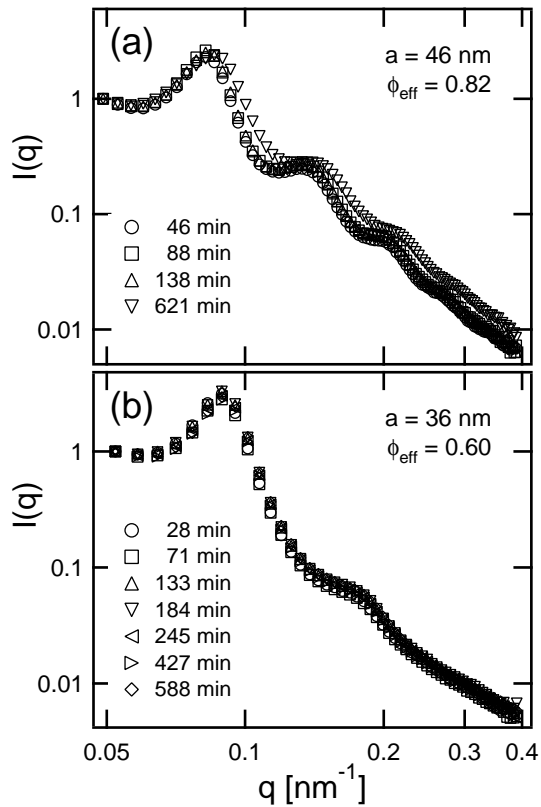}
\caption{X-ray scattering intensities for nanoemulsions of (a) droplet radius 46 nm (set A) with $\phi_{eff}=0.82$ and (b) droplet radius 36 nm (set B) with $\phi_{eff}=0.60$ measured at several waiting times $t_w$ following the loading of the nanoemulsions into sample cells, as specified in the figure legends.  The intensities are normalized with respect to the intensity at the smallest wave vector.  The peaks in $I(q)$ below $q = 0.10$ nm$^{-1}$ correspond to the first interparticle structure factor peak.  Additional correlation peaks are also observable at higher wave vector. }
\label{fig1}
\end{figure}
%%%%%%%%%%%%%%%%%%%%%%%%%%%

Figures 2(a) and 2(b) display results for the corresponding intensity autocorrelation function $g_2(q,t)$  measured at a wave vector near the first peak in $I(q)$ for the nanoemulsions from set A with $\phi_{eff}=0.82$ and set B with $\phi_{eff} = 0.60$, respectively, at the same waiting times as in Figs.~1(a) and 1(b).  The correlation functions decay on a time scale that increases with increasing $t_w$ immediately following the loading of the nanoemulsions into the sample cells, but reach steady-state behavior at larger $t_w$.  The dynamic structure factor $f(q,t)$ can be calculated from $g_2(q, t)$  using the Siegert relation~\cite{lumma_pre}.  Modeling $f(q,t)$ with a stretched exponential lineshape $f(q,t) = f_qexp[(-t/\tau)^\beta]$, the intensity autocorrelation function has the form 
\begin{equation}\label{g2Eq}
 g_2(q, t) =  1+b[f_qexp[-(t/\tau)^{\beta}]]^2
\end{equation}
\noindent where $f_q$ is the short-time ($t < $ 1 s) plateau amplitude of $f(q,t)$, $\tau$ is the terminal relaxation time, $\beta$ is the stretching exponent that characterizes the shape of the correlation function, and  $b \simeq$ 0.30 is the Siegert factor determined from measurements on a static aerogel sample.  For all the nanoemulsions and all wave vectors, fits to Eq.~(\ref{g2Eq}) agree closely with the data except at very large $t$ and small $g_2(q,t)$.  The lines in Fig.~2 are the results of fits to Eq.~(\ref{g2Eq}) for $g_2(q,t) > 1.03$, illustrating the good agreement at small $t$ and a crossover to a slower decay in the measured $g_2(q,t)$ at larger $t$.

%%%%%%%%%%%Figure 2 Here%%%%%%%%%
\begin{figure}
\includegraphics[scale=1.0]{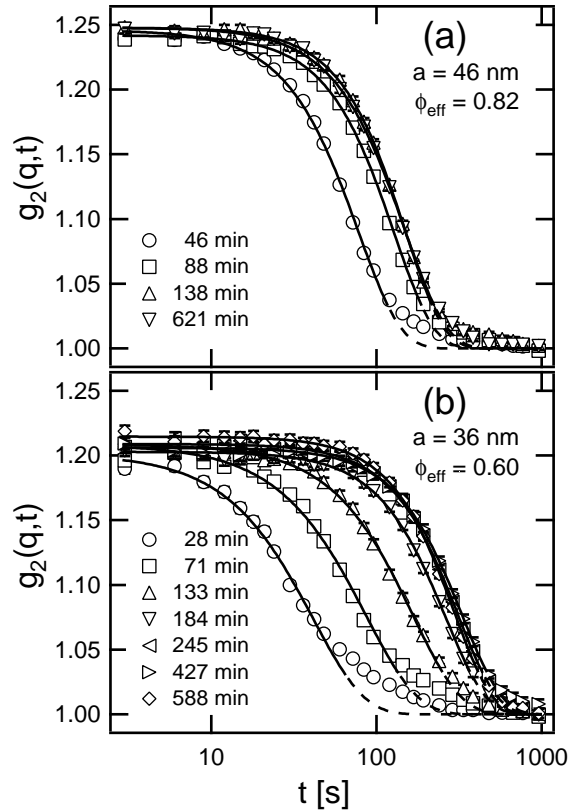}
\caption{Intensity autocorrelation functions for nanoemulsions of (a) droplet radius 46 nm (set A) with $\phi_{eff}=0.82$ at $q=0.08$ nm$^{-1}$  and (b) droplet radius 36 nm (set B) with $\phi_{eff}=0.60$ at $q=0.10$ nm$^{-1}$ measured several waiting times $t_w$ following the loading of the nanoemulsions into sample cells.  Symbols refer to the same waiting times as in Fig.~1.  The solid lines are the results of fits to Eq.~(1) over the range $g_{2}(q,t) >1.03$ and are extrapolated with dashed lines beyond this range to illustrate the deviations from Eq.~(1) at large $t$.}
\label{fig2}
\end{figure}
%%%%%%%%%%%%%%%%%%%%%%%%%%%

\subsection{Fast Dynamics}

Results for the short-time plateau value of $f(q,t)$ extracted from fits to Eq.~(1) are consistently less than one, as illustrated in Fig.~3(a), which displays values of $f_q$ as a function of wave vector for nanoemulsions from set A with $\phi_{eff} =$ 0.62 and 0.82.  The observation that $f_q < 1$ indicates a partial decay of the dynamic structure factor at inaccessibly short times, suggesting a two-step relaxation in the dynamics.  Two-step relaxations are a signature feature of glassy particle dynamics.  In colloidal suspensions near the glass transition, the particle dynamics separate into a fast localized component in which a particle moves within the restricted volume defined by the positions of its neighbors, creating a partial decay in $f(q,t)$, and slow component in which the particle diffuses away from its center-of-mass position with cooperative motion from its neighbors, leading to the final decay of $f(q,t)$.  We thus interpret the fast partial decay of $f(q,t)$ of the nanoemulsions in terms of localized motion of the droplets within the constraints of their neighbors.  A similar conclusion was reached regarding fast dynamics in concentrated emulsions of micrometer-scale droplets based on dynamic light scattering measurements~\cite{gang}.  However, for both the emulsions of micrometer-scale droplets and colloidal suspensions, $f_q$ displays a pronounced peak at a wave vector near the position of the first peak in the interparticle structure factor $S(q)$~\cite{gang,beck}.  In contrast, $f_q$ for the nanoemulsions shown in Fig.~3(a) is largely featureless as a function of $q$.   Mode coupling theory, which provides a framework for the dynamics in colloidal suspensions near the glass transition, predicts a peak in the amplitude of $f_q$ near the interparticle structure factor peak~\cite{mct}.  However, the magnitude of the peak predicted by the theory depends in detail on the interparticle potential and the resulting form of $S(q)$~\cite{beck}.  The relatively featureless $f_q$ for the concentrated nanoemulsions could thus reflect the distinct interparticle interactions between the charged, deformable droplets and the interparticle structure factor that has a relatively weak correlation peak as compared with that of hard-sphere glasses~\cite{graves,tomJPCB}.  Alternatively, the wave-vector independence of $f_q$ could indicate that the fast dynamics in the nanoemulsions are predominately local rotational motions that potentially involve one or more droplets. 
    
As with wave vector, $f_q$ does not show any clear variation with $t_w$, consistent with an unchanging structure; however, it does vary systematically with droplet volume fraction.  The dashed lines in Fig.~3(a), representing the averages of $f_q$ for the two volume fractions, illustrate this variation.  Figure 3(b) shows the average plateau value $\bar{f_q}$ for the nanoemulsions in sets A and B.  To obtain $\bar{f_q}$, the values of $f_q$ are averaged over both wave vector and $t_w$.  The resulting trends, in which the short time plateau increases with increasing droplet volume fraction, indicate that the range of the localized motion becomes smaller with increasing volume fraction, as expected for repulsively interacting droplets.

%%%%%%%%%%%Figure 3 Here%%%%%%%%%
\begin{figure}
\includegraphics[scale=1.0]{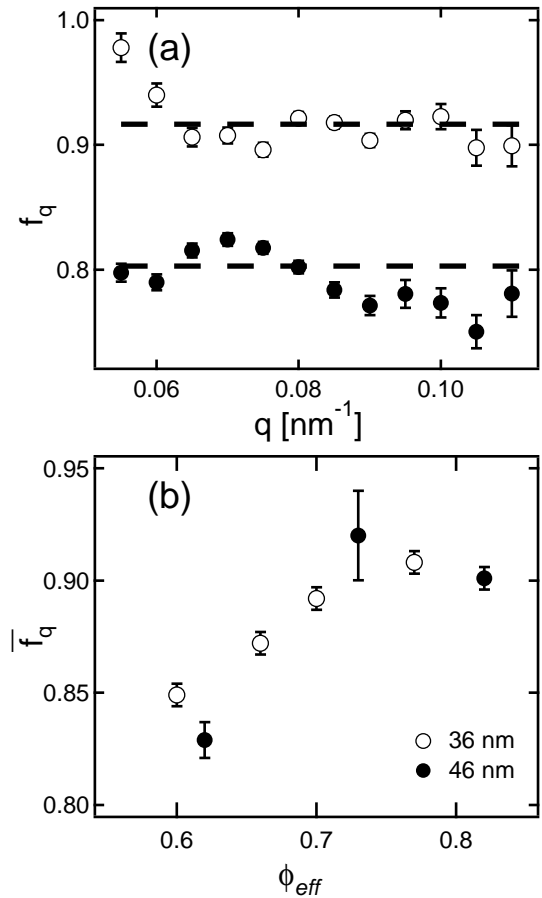}
\caption{(a) Short-time plateau value of ${f(q,t)}$ versus wave vector for nanoemulsions with droplet radius 46 nm and volume fractions $\phi_{eff}=0.82$ (open) and 0.62 (filled) measured at $t_w =$ 130 minutes.  The dashed lines show the average values for each volume fraction at this waiting time.  (b) Short-time plateau values averaged over $q$ and $t_w$ as a function of droplet volume fraction for sets A (filled) and B (open).  }
\label{fig3}
\end{figure}
%%%%%%%%%%%%%%%%%%%%%%%%%%%

\subsection{Slow, Non-Diffusive Dynamics}

While the fast, localized component of the droplet motion in the nanoemulsions is common to colloidal glasses, in contrast, the slow dynamics characterized by the decay in $g_2(q,t)$ are clearly distinct from the cooperative diffusion in colloidal suspensions near the glass transition.  Instead, the shape of $g_2(q,t)$ and its wave vector dependence exemplify two salient features of non-diffusive dynamics that are apparently universal to a range of disordered soft solids.  First, fits to Eq.~(1) give $\beta = 1.7 \pm 0.2$, implying a compressed exponential correlation function.  This lineshape, which we observe for all of the nanoemulsions, displays no systematic variation with $t_w$ or $\phi_{eff}$, although some results suggest that $\beta$ decreases slightly with increasing $q$.  Second, the relaxation time $\tau$ extracted from the fits varies with wave vector as  $\tau \sim q^{-z}$ with $z \approx 1.0$, as shown in Fig.~4 for the nanoemulsion from set A with $\phi_{eff} = 0.82$.  Similar compressed exponential lineshapes with  $\tau \sim q^{-1}$ have been observed previously in dynamic light scattering and XPCS studies of colloidal gels~\cite{cipelletti,chung}, clay suspensions~\cite{bandyopadhyay,bellour}, dense ferrofluids~\cite{robert}, concentrated emulsions of micrometer-scale droplets~\cite{cipelletti-farad}, and a block copolymer mesophase~\cite{falus}.  Typically, for these other disordered soft solids $\beta \approx 1.5$, slightly smaller than we observe for the nanoemulsions.  Regardless of the precise value, however, such compressed correlation functions with $\beta >1$ and $\tau \sim q^{-1}$ are inconsistent with purely diffusive particle motion.  Instead, as discussed previously~\cite{cipelletti-farad}, they indicate ballistic motion with a broad distribution of velocities and a characteristic velocity, $v_0 \equiv 1/(\tau q)$.  Based on a heuristic argument by Cipelletti {\it et al.}~\cite{cipelletti,cipelletti-farad}, Bouchaud and Pitard~\cite{pitard} have introduced a microscopic model for these ballistic dynamics, describing them in terms of elastic strain deformation in response to heterogeneous local stress.  Specifically, Bouchaud and Pitard picture the source of this stress as a random population of particle rearrangement events that create point-like dipolar stress fields whose intensities $P(t)$ grow linearly over a period of time $\theta$ to a maximum $P_0$, and they show that the resulting strain leads to a compressed exponential lineshape and  $\tau \sim q^{-1}$~\cite{pitard}.  More recently, Duri and Cipelletti have shown that, at least for dilute colloidal gels, the strain grows linearly in time only in a time-averaged sense and the dynamics are more accurately described in terms of temporally heterogeneous, discrete displacements~\cite{duri}.  An interesting question is whether such temporally heterogeneous motion also describes the strain in concentrated soft solids like nanoemulsions that display this characteristic non-diffusive dynamics.

%%%%%%%%%%%Figure 4 Here%%%%%%%%%
\begin{figure}
\includegraphics[scale=1.0]{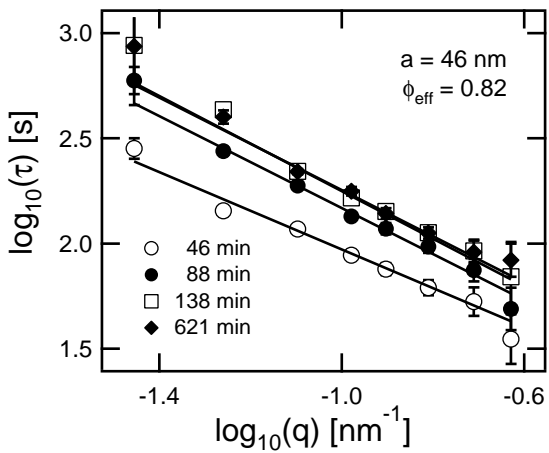}
\caption{Characteristic relaxation time $\tau$ as a function of a wave vector $q$ for a nanoemulsion with droplet radius 46 nm (set A) and $\phi_{eff}=0.82$ measured at several waiting times, as specified in the figure legend.  The solid lines are the results of fits to the form $\tau \sim q^{-z}$, which give an exponent $z$ ranging from 0.87 to 1.10 with an average $z = 1.03 \pm 0.09$. }
\label{fig4}
\end{figure}
%%%%%%%%%%%%%%%%%%%%%%%%%%%

A key parameter in the Bouchaud-Pitard model is the time scale $\tau_q=\frac{4\pi K^2\theta^2}{P_0\gamma q}$, where $K$ is the elastic modulus of the material and $\gamma$ is a friction coefficient related to dampening of the strain motion.  For $t << \tau_q$, the model predicts $\beta = 1.5$, as observed in most previous systems displaying these non-diffusive slow dynamics.  However,  if $t/\tau_q$ is not very small, an effective exponent somewhat larger than 1.5 is expected.  Further, for $t >> \tau_q$ a new regime of behavior dominates in which $\beta = 1.25$.   Thus, the observed lineshapes, like those displayed in Fig.~2, which follow a compressed exponential form with $\beta \approx 1.7$ at smaller $t$ and a more stretched decay at larger $t$, are qualitatively consistent with the idea that the measurements on the nanoemulsions probe a range of dynamics in which $t \gtrsim \tau_q$.  However, the considerable scatter in the data at very large $t$ (where $g_2(q,t) \lesssim 1.03$ and deviations from the fits to Eq.~(1) in Fig. 2 are apparent) does not permit a quantitative analysis of the lineshape in this regime.

\subsubsection{Relaxation Times}

Figure 5(a) displays the relaxation time $\tau$ as a function of $t_w$ for nanoemulsions in set A with three different droplet volume fractions measured at a wave vector near the first peak in $I(q)$.  For two effective volume fractions, $\phi_{eff} =$ 0.62 and 0.82, the relaxation time grows rapidly at short $t_w$ and saturates to a steady-state value at longer waiting times.  Similar rapid increases in $\tau$ followed by steady-state behavior are also observed with the nanoemulsions in set B, as shown in Fig.~5(b).  However, the magnitude of the change in $\tau$ at early $t_w$ varies considerably among different nanoemulsions.  For example, as shown in Fig.~5(a), the nanoemulsion with $\phi_{eff}=0.73$ in set A displays no appreciable increase in $\tau$ over the measured range of $t_w$, suggesting that the first measurement of $g_2(q,t)$ on this nanoemulsion was performed at too late a waiting time to capture any transient behavior.

%%%%%%%%%%%Figure 5 Here%%%%%%%%%
\begin{figure}
\includegraphics[scale=1.0]{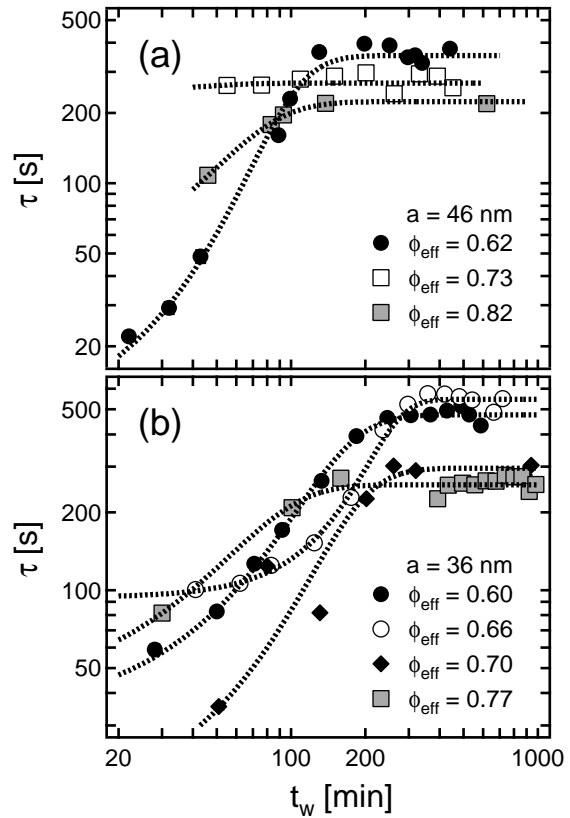}
\caption{Characteristic relaxation time as a function of waiting time for nanoemulsions with (a) droplet radius 46 nm and (b) droplet radius 36 nm with various droplet volume fractions, specified in the figure legends, measured at a wave vector near the first peak in $I(q)$  ($q = 0.08$ nm$^{-1}$ for (a) and $q = 0.10$ nm$^{-1}$ for (b)).  The dashed lines are the results of fits to Eq.~(3), which implies two components to the characteristic strain velocity.}
\label{fig5}
\end{figure}
%%%%%%%%%%%%%%%%%%%%%%%%%%%

 To model the evolution of $\tau$, we suppose that the characteristic strain velocity has two components, one that decays exponentially with increasing $t_w$ and one that is independent of waiting time, 
\begin{equation}\label{velocityEq}
 v_0 = (v_i - v_s)exp(-t_w/\Gamma)+v_s,
\end{equation}
where $v_i$ is the characteristic velocity at $t_w = 0$, $\Gamma$ is the decay time of the transient component, and $v_s$ is the steady-state characteristic velocity reached at $ t_w >> \Gamma$.  The corresponding relaxation time is therefore given by
\begin{equation}\label{tauEq} 
\tau = (((v_i-v_s)exp(-t_w/\Gamma)+v_s)q)^{-1},
\end{equation}
The dashed lines in Figs.~5(a) and 5(b) are the results of fits to Eq.~(\ref{tauEq}).  For the nanoemulsions in which the measurements capture an appreciable increase in $\tau$ at early $t_w$, Eq.~(\ref{tauEq}) describes the evolution accurately.  The time scale for the decay of the transient strain motion is found to be in the range $\Gamma \approx 10 - 20$ minutes for all cases.  For significantly larger waiting times, $t_w >> \Gamma$, this transient component contributes negligibly, and the strain motion is dominated by the steady-state component.

\subsubsection{Transient Component}

The presence of two components to the strain motion indicates two distinct sources of local stress relaxation.  The transient motion, which dominates at small $t_w$, has a characteristic velocity that decays rapidly with increasing $t_w$ leading to the corresponding exponential growth in $\tau$.  A similar exponential growth in $\tau$ at early $t_w$ has been observed in a number of disordered soft solids including colloidal gels~\cite{cipelletti}, clay suspensions~\cite{bellour}, and dense ferrofluids~\cite{robert}, indicating that such an evolution in these non-diffusive dynamics is fairly generic.  This rapid evolution typically follows a significant perturbation of the system, such as a quench from fluid to disordered solid or a large mechanical stress, and these dynamics thus likely reflect a relaxation from the state created by the perturbation.  Thus, we associate the transient strain motion in the nanoemulsions with residual stress that is introduced by loading the nanoemulsions into the sample cells and that relaxes through slow droplet rearrangements.   The extent of mechanical perturbation associated with the sample loading process was not quantitatively controlled in the experiments, and, therefore, the amount of residual stress introduced presumably varied among the samples.  The wide range in the amount that $\tau$ changes at early $t_w$ seen in Fig.~5 likely reflects this variability.   In the context of stress dipoles, the perturbation thus creates many dipoles at the same formation time, $t_w = 0$, and the subsequent dynamics are dominated by the growth in intensity of these dipoles.  We note this scenario is different from the one considered by Bouchaud and Pitard, whose model assumes a steady-state formation rate $\rho$ of new stress dipoles.  (Although, the model also includes a mechanism for the dynamics to evolve, or ``age'', through an age-dependent $\rho$ that leads to a power-law growth in $\tau$ with age.)   One question therefore is whether the addition of a large population of stress dipoles formed at $t_w=0$ into the Bouchaud-Pitard model could capture the exponential growth in $\tau$ versus $t_w$ observed experimentally.  Presumably, to obtain the observed behavior the model would need to consider a distribution of growth periods $\theta$ of these dipoles, rather than a single characteristic period.  

An interesting comparison can be made between these transient droplet dynamics in concentrated nanoemulsions observed with XPCS and those of the bubbles of macroscopic foam studied with diffusing wave spectroscopy (DWS).  In the foam experiments, $\tau$ is large immediately following the cessation of shear and decreases exponentially with waiting time as the system recovers~\cite{gopal,cohen-addad}.  Thus, a waiting-time dependence essentially opposite to that in Fig.~5 is observed.  This contrasting behavior presumably reflects both the differing mechanical perturbation imposed in the foam studies as compared with this study as well as the differing compliances of the two systems and suggests an analogy with the ``over-aging'' versus ``rejuvenation'' scenario that occurs in the shear response of glasses~\cite{viasnoff,lacks,joos,isner}.  Specifically, the large correlation time immediately following the shearing of the foam is interpreted as a consequence of homogenizing the internal strain field~\cite{gopal}.  That is, the shear assists the system in reaching a lower-energy configuration much like the phenomenon envisioned in ``over-aging''.  In contrast, the perturbation of the nanoemulsions involved in loading them in the sample cells has the opposite ``rejuvenating'' effect of imparting energy into the system in the form of local regions of poorly packed droplets, leading to an initially dense population of growing stress dipoles.  Future experiments on nanoemulsions that investigate the effects of shear flow over a wide range of shear rates might capture the transition between these classes of response, and could provide further insight into the effects of shear on soft glassy materials.

%%%%%%%%%%%Figure 6 Here%%%%%%%%%
\begin{figure}
\includegraphics[scale=1.0]{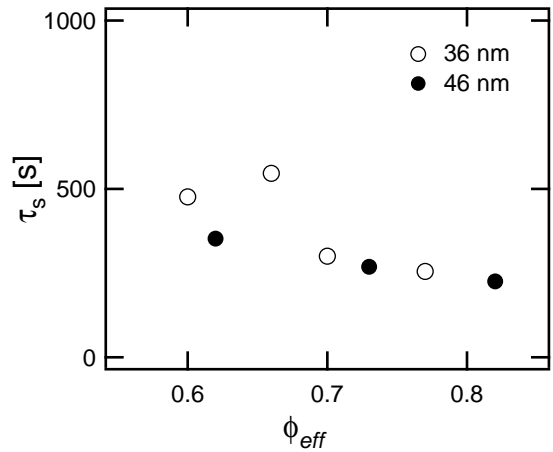}
\caption{Steady-state characteristic relaxation time measured at large waiting time $t_w$ for nanoemulsions with droplet radii $a = $ 46 nm (solid) and $a = $ 36 nm (open) as a function of effective volume fraction.  The relaxation times correspond to wave vectors near the first peak in $I(q)$  ($q = 0.08$ nm$^{-1}$ for $a = $ 46 nm and $q = 0.10$ nm$^{-1}$ for $a = $ 36 nm).}
\label{fig7}
\end{figure}
%%%%%%%%%%%%%%%%%%%%%%%%%%%

\subsubsection{Steady-State Component}

As mentioned above, the values of $\tau$ that are independent of waiting time at large $t_w$ in the nanoemulsions indicate a second, steady-state source of local stress-relaxation events that is distinct from the residual stress of sample loading.  The fact that $\tau$ is independent of waiting time further suggests a means for exploring these dynamics systematically.  Specifically, while such non-diffusive dynamics have been identified in a host of disordered soft-solid materials, the significance of this strain motion for other material properties is unclear.  The seeming insensitivity of the steady-state component  in nanoemulsions to sample history should simplify the study of its dependence on other material parameters, and clarifying their relation to other parameters might shed light on the role of these dynamics in dictating material behavior.   As a first investigation into the nature of the steady-state dynamics, we have characterized the dependence of the strain motion on droplet volume fraction.  Figure 6 shows the steady-state relaxation time $\tau_s \equiv 1/(v_sq)$, measured for $t_w >> \Gamma$, as a function of $\phi_{eff}$ for nanoemulsion sets A and B.   Over the ranges of $\phi_{eff}$ investigated, the dynamics are only weakly sensitive to volume fraction, with the steady-state relaxation time increasing by roughly a factor of two with decreasing $\phi_{eff}$.  As mentioned above, rheology studies have shown that the shear modulus decreases by more than a factor of 100 with decreasing $\phi_{eff}$ over these ranges~\cite{rheo}.  

At first glance, the relatively weak dependence of the strain motion on $\phi_{eff}$ is thus surprising since the Bouchaud-Pitard model predicts that $\tau \sim \frac{K\theta^{1/3}}{\rho^{2/3}P_0q}$, where $K$ is an elastic constant~\cite{pitard}.  (In the model $K$ represents the compression modulus, so this comparison is not exact; however, a more refined model that includes consideration of the shear modulus would differ only in numerical factors~\cite{pitard}.)  However, the variation in modulus with volume fraction occurs through the increase in the droplet-droplet repulsion potential with increasing $\phi_{eff}$, and a change in the interaction potential should also affect $P_0$, the strength of the stress dipoles driving the strain motion.   If $K$ and $P_0$ possess the same dependence on interaction potential, then the effects on the strain motion of changing $\phi_{eff}$ should cancel since $\tau \sim K/P_0$, consistent with our observation that $\tau_s$ varies only weakly with $\phi_{eff}$.  Additional XPCS measurements on nanoemulsions with varying concentrations of NaCl added to the continuous phase support this picture. The salt, added to nanoemulsions from set A with $\phi_{eff} =0.73$ to concentrations as large as $c = 160$ mM, decreases significantly the shear modulus by screening the Coulombic interaction between droplets.   However, XPCS measurements on the nanoemulsions with added salt revealed that the steady-state strain dynamics are essentially unchanged. 

While the weak dependence of the non-diffusive dynamics on $\phi_{eff}$ and added salt provides some insight into the mechanisms driving the strain motion, the microscopic origin of the local sources of steady-state stress in the nanoemulsions remains a question.  One possibility is droplet coarsening, which is known to drive the dynamics in macroscopic foams observed with DWS~\cite{durian1,durian2}.  However, the correlation function observed for the foams with DWS is quite distinct from the compressed exponential lineshape measured with XPCS on the nanoemulsions, indicating quite different microscopic dynamics.  Further, the long shelf life of the nanoemulsions indicates that any coarsening proceeds at an extremely slow rate, making it an unlikely candidate for creating sources of stress.  
However, we do note that some of the nanoemulsions did display small variations in $I(q)$ during the course of the measurements, as illustrated by Fig.~1(a), which might indicate that the emulsions are made less stable by the x-ray measurement itself.  Thus, the sources of stress might originate as a consequence of the measurement.  For example, contact between the nanoemulsions and the sample cell could cause droplet coalescence and such coalescence events could drive the strain response.  However, we do not expect that the materials in contact with the nanoemulsions -- specifically, stainless steel, polyimide, and epoxy -- would be problematic.  Also, analysis of the measurements to discriminate dynamics along different wave-vector directions indicates that the dynamics are isotropic, suggesting that the sources of stress are positioned randomly and are not restricted to the surfaces.  Radiation damage is another potential source of local stress introduced by the measurement.  However, efforts to identify evidence of radiation damage by varying exposure times were negative, and the protocol of making each measurement of $g_2(q,t)$ on a new region of sample should have prevented any cumulative damage.   A future experiment that might provide insight would be a study in which droplet coarsening is intentionally accelerated.  One approach might be to investigate the temperature dependence of the dynamics to see if the strain motion reflects the thermally activated behavior expected for coarsening.  Another would be to employ nanoemulsions comprised of oils that are less immiscible in water, so that the stability against coarsening is compromised.  Testing whether such a change alters qualitatively the droplet-scale dynamics or merely increases the characteristic velocity of the observed strain motion would help elucidate the possible role of coarsening in the current study. 

Another possible source of the steady-state, non-diffusive dynamics is local, irreversible shear deformation that results from thermal expansion in response to temperature fluctuations.  In a recent microscopy study of concentrated multilamellar vesicles, Mazoyer {\it et al.}~\cite{mazoyer} observed slow dynamics corresponding to such deformation and suggested that they might be relevant to the slow, non-diffusive dynamics seen in light scattering and XPCS studies of soft glassy materials.  While such irreversible shear deformation represents a physically appealing microscopic origin for the steady-state motion, we note that the temperature control in the XPCS experiments limited the fluctuations in sample temperature to approximately $\pm 0.001$ K.  This fluctuation amplitude was significantly smaller than that in the microscopy study, raising doubt as to whether thermal expansion fluctuations would be large enough in the nanoemulsions to drive such shear deformation.   An additional, related issue in the XPCS measurements is the possibility of local heating due to x-ray absorption.  Calculations and experiments on materials sensitive to this effect indicate that the sample temperature within the 20 $\mu$m $\times$ 20 $\mu$m cross section of the beam increased by approximately 0.004 K due to beam heating.  This temperature increase occurred with the periodicity of the x-ray exposures, once every 3 s, and so could conceivably play a role similar to that of the temperature fluctuations in Mazoyer {\it et al.}  However, since the amplitude of the increase was so small, we believe the thermal expansion caused by this fluctuation again was likely to be too small to drive the observed dynamics.  A future experiment that might elucidate the potential role of thermal fluctuations would be to introduce temperature oscillations of varying amplitude and frequency to test whether they systematically influence the steady-state non-diffusive dynamics.

\section{Conclusion}

In conclusion, we have found that the slow dynamics of nanoemulsions probed with XPCS are characterized by non-diffusive motion similar to that observed in a number of disordered soft solids.  The compressed exponential form for $f(q,t)$ and inverse relationship between relaxation time and wave vector that are the signatures of these dynamics match well to the predictions of the model introduced by Cipelletti {\it et al.}~\cite{cipelletti,cipelletti-farad} and developed by Bouchaud and Pitard~\cite{pitard} in which this motion corresponds to strain from heterogeneous, local stress.  However, the large range of disordered soft solids in which these dynamics have been observed suggests that a more general principle might underlie the formation of such sites of stress (perhaps one tied to the mechanical properties of systems undergoing a jamming transition~\cite{jamming}) and calls for further investigation into the origin of these dynamics and their role in determining material properties.  The relatively unique aspect of the dynamics in the nanoemulsions, which is also shared by the recently reported dynamics in a  polymer-based sponge phase~\cite{falus}, is their steady-state behavior.  This apparent lack of evolution in the stress relaxation has a number of significant implications.  First, it highlights an observation made previously~\cite{chung} that  these dynamics are distinct from traditional aging seen below the glass transition in hard disordered solids such as molecular liquids and polymer melts~\cite{struik,leheny,lunkenheimer}.  Second, it opens the possibility for systematically tuning the strain motion by changing material parameters.  This latter feature makes the nanoemulsions a particularly appealing model system for future investigations into the nature of these slow, non-diffusive dynamics.

\vspace{1 cm}
\noindent
{\bf Acknowledgements:}
We thank R. Bandyopadhyay, D. Durian, and S. Mochrie for helpful discussions and S. Narayanan and A. Sandy for their assistance.  Funding was provided by the NSF (DMR-0134377).  Use of the APS was supported by the DOE, Office of Basic Energy Sciences, under Contract No.~W-31-109-Eng-38.

\end{document}